\UseRawInputEncoding

\documentclass[aps,prl,twocolumn,amsmath,amssymb,nofootinbib,superscriptaddress,floatfix]{revtex4-1}

\usepackage[utf8]{inputenc}
\usepackage{gensymb}
\usepackage{amsmath}
\usepackage{amssymb}
\usepackage{amsthm}
\usepackage{mathtools}
\usepackage{mathdots}
\usepackage{amsfonts}
\usepackage{bbm}
\usepackage{xr}
\usepackage{bbold}
\usepackage{newpxtext}
\usepackage{epsfig,color,dsfont,upgreek,physics}
\usepackage{mathrsfs}
\usepackage{multirow}
\usepackage[makeroom]{cancel}
\usepackage{mathtools}

\usepackage{graphicx}
\usepackage{dcolumn} 
\usepackage{bm} 
\usepackage{float} 
\usepackage[driverfallback=dvipdfm]{hyperref} 
\usepackage{longtable}
\usepackage{ulem}   
\allowdisplaybreaks

\newcommand{\bs}[1]{{\boldsymbol{#1}}}

\begin{document}

 \title{Exciton-Anyon Binding in Fractional Chern Insulators: Spectral Fingerprints
}

\author{Tianhong Lu}
\affiliation
{
Department  of  Physics,  Emory  University,  400 Dowman Drive, Atlanta,  GA  30322,  USA
}
\author{Luiz H. Santos}
\affiliation
{
Department  of  Physics,  Emory  University,  400 Dowman Drive, Atlanta,  GA  30322,  USA
}

\begin{abstract}
Transition--metal dichalcogenides (TMDs) uniquely combine topological electronic states realized without external magnetic fields with a strong optical response arising from long--lived excitons. Motivated by this confluence, we investigate an interacting fermion--boson system formed by coupling an exciton to a quasihole of a fractional Chern insulator (FCI) at filling fraction \(1/3\). We introduce a kagome--lattice fermion--boson model hosting an electronic FCI and a mobile exciton whose dispersion is tunable from a parabolic band to a flatband. Using exact diagonalization, we demonstrate the emergence of exciton--quasihole bound states controlled by the repulsive electron--exciton interaction \(V_{\mathrm{FB}}\) and the exciton kinetic energy \(t_{\mathrm{B}}\). These states appear as low--lying levels in the fermion--boson spectrum, well separated from the scattering continuum, and arise despite repulsive interactions due to a residual attraction to the local charge depletion associated with a quasihole. Reducing \(t_{\mathrm{B}}\) enhances this effect by favoring interaction--dominated binding. Our results provide a model description of moir\'e TMD heterostructures, including fractional Chern insulating twisted bilayer MoTe$_2$ proximitized by excitonic TMD heterobilayers, where we estimate exciton--quasihole binding energy scales of {\(0.8\)--\(1.2\)~meV}, placing these effects within reach of photoluminescence spectroscopy.
\end{abstract}
\date{\today}

\maketitle

\begin{figure}[!t]
    \centering
    \includegraphics[width = \linewidth]{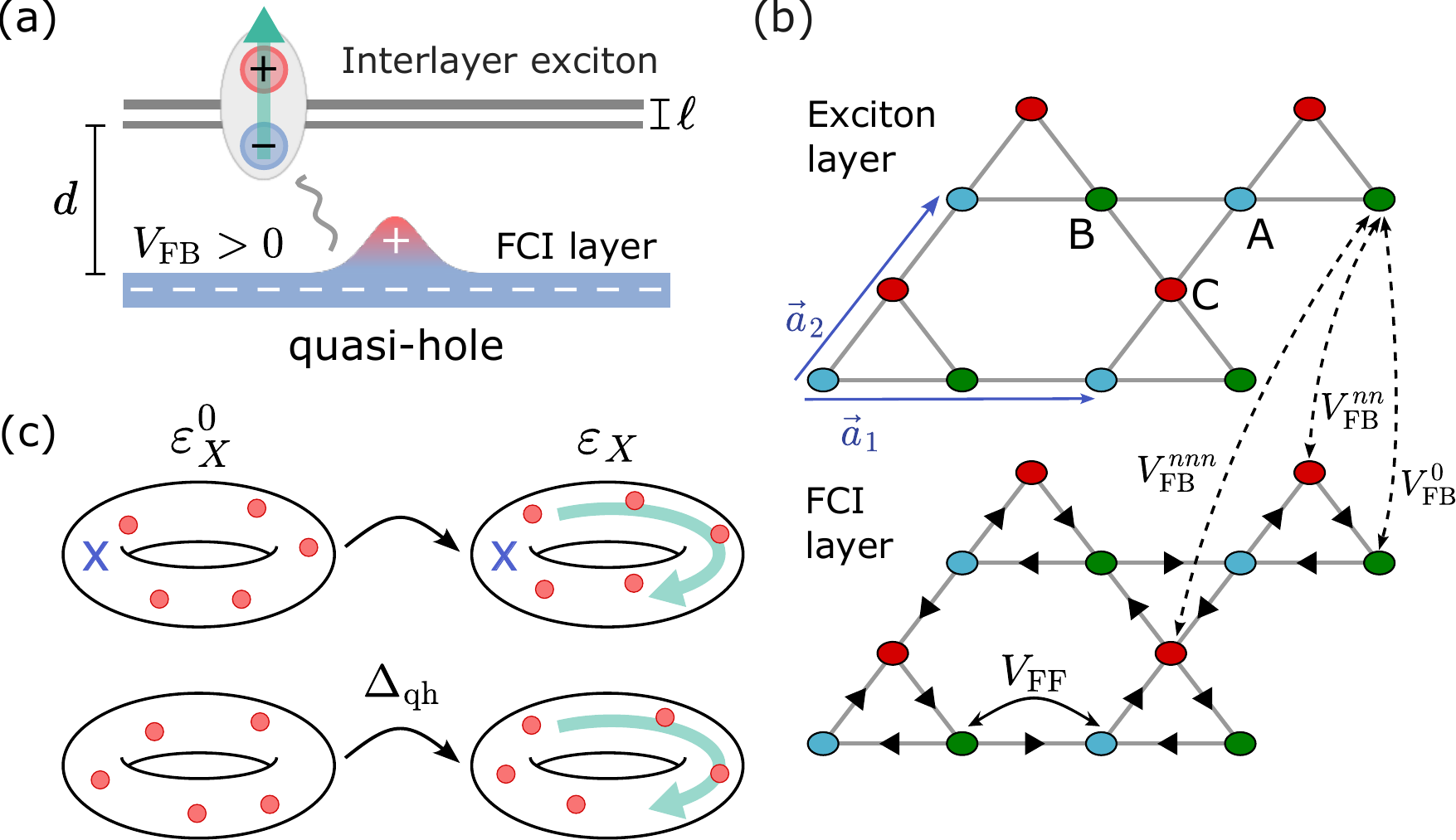}
\caption{The exciton--FCI interacting system. 
(a) Schematic of an interlayer exciton in proximity to an FCI layer. The exciton dipole orientation induces repulsive interactions with the electronic background while generating a residual attraction to a quasihole.  
(b) Effective lattice model consisting of superposed kagome lattices for the FCI and exciton layers. Red, green, and blue dots denote sublattices A, B, and C, respectively. Arrows indicate fermionic hopping phases \(e^{i\phi}\), with \(\phi=\pi/4\).  
(c) Setup for characterizing exciton--quasihole binding, with binding potential (top) and the quasihole excitation gap (bottom). Excitons and electrons are represented by X symbols and red dots, respectively, while the arrow denotes a flux insertion on the torus that nucleates a quasihole.}
\label{fig: kagome+cartoon}
\end{figure}

\noindent
\textit{Introduction--}
A hallmark of strongly correlated quantum matter is the emergence of topological order from the interplay of interactions and topology \cite{wen2004quantum}, accompanied by anyonic excitations with fractional charge \cite{laughlin1983anomalous} and Abelian fractional statistics \cite{halperin1984statistics,arovas1984fractional}. 
{Such phases have now been realized as fractional Chern insulators (FCIs) in moir\'e transition-metal dichalcogenides (TMDs) \cite{cai2023signatures,zeng2023thermodynamic,park2023observation,xu2023observation} and multilayer graphene \cite{lu2024fractional} without external magnetic fields -- and TMDs uniquely combine these topological states with strong optical response and robust excitons \cite{wang2018colloquium,regan2022emerging}, setting the stage for exploring fractionalized matter coupled to excitons.} 
In this Letter, we characterize bound states formed by coupling an exciton -- harbored in a TMD heterobilayer in proximity to a fractional Chern insulator -- to an Abelian quasihole, as shown in Fig.~\ref{fig: kagome+cartoon}~(a).
{
Within FCI and FQH plateaus, the bulk elementary charged excitations are emergent quasiparticles with fractional charge and anyonic statistics, localized by disorder~\cite{halperin1982quantized,
nakamura2020direct,nakamura2023fabry} at an energy cost below the charge
gap~\cite{laughlin1983anomalous,halperin1984statistics};
the exciton thus acts as a local optical probe of these fractionalized charges, with
the binding energies we compute providing direct spectral access to the quasihole sector.}
{We present strong numerical evidence that lattice fractionalized phases with nonuniform quantum geometry and strong lattice effects host a rich structure of exciton--quasihole bound states, extending exciton--anyon binding beyond FQH systems \cite{mostaan2025anyon,wagner2025sensing}.}
These hybrid excitations offer a controlled setting to explore fractionalization beyond the electronic sector and allow optical spectroscopy to directly access anyonic matter \cite{cai2023signatures,park2023observation,hayakawa2013real}. 

{Despite the exponential growth of the fermion--boson Hilbert space and absence of a Landau--level description, we identify a minimal, translationally invariant fermion--boson kagome lattice, illustrated in Fig.~\ref{fig: kagome+cartoon}(b), with three key ingredients:
First, a $1/3$-filled FCI 
stabilized by repulsive interactions
in an isolated $C=1$ Chern band with a well-defined quasihole manifold.
Second, a bosonic sector tunable between parabolic and flat dispersions via the hopping $t_\text{B}$.
Third, a local repulsive fermion--boson coupling $V_{\text{FB}}$, modeling an interlayer exciton with electric dipole oriented away from the proximate FCI layer, shown in Fig.~\ref{fig: kagome+cartoon}(a).
This system provides an effective model for heterostructures such as twisted bilayer MoTe$_2$ hosting an FCI in proximity to a TMD heterobilayer supporting interlayer excitons, including MoSe$_2$/WSe$_2$ \cite{tran2019evidence,kunstmann2018momentumspace,alexeev2019resonantly,forg2021moire} and WSe$_2$/WS$_2$ \cite{xiong2023correlated,park2023dipole,gao2024excitonic,lian2024valley,wang2023intercell}.} 

{Using exact diagonalization on periodic lattices with near--unity aspect ratio, we identify exciton--quasihole bound states by comparing spectra with and without an exciton upon quasihole creation, shown in Fig.~\ref{fig: kagome+cartoon}(c). Despite the repulsive fermion--boson interaction, a residual attraction to the quasihole charge depletion gives rise to bound states emerging between the spectral minimum and the onset of the quasihole continuum -- establishing a route to optically probe anyonic matter in lattice platforms where fractionalization and bosonic excitations coexist without magnetic fields.}

\begin{figure}[!t]
    \centering
    \includegraphics[width =\linewidth]{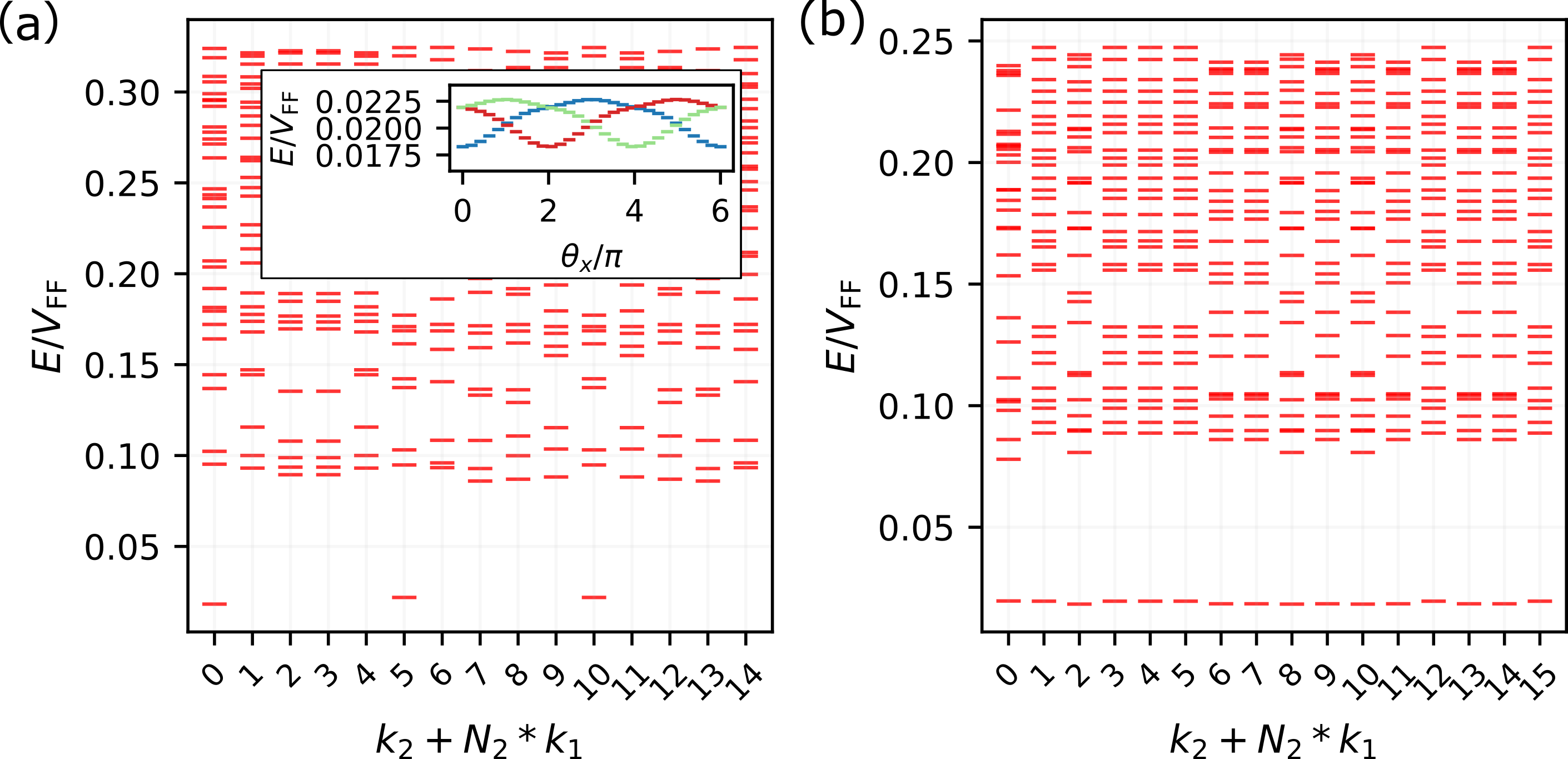}
\caption{(a) FCI spectra at 1/3 filling for the 3x5 kagome lattice. The fermion-fermion interaction strength is set as constant $V_{\text{FF}}=2.5$.  Inset shows the spectral flow of the 3-fold degenerate 1/3 FCI states when varying $\theta_x$. (b) The corresponding quasihole spectrum by adding one flux to the 1/3 FCI system. The aspect ratios chosen here render the lattices as isotropic as possible to avoid strong finite-size artifacts.}
\label{fig: FCI spectrum}
\end{figure} 

\noindent
\textit{Fermi-Bose Model --}
We consider spinless fermions described by the nearest-neighbor tight-binding Hamiltonian $H_{f} = \sum_{\langle i j \rangle} t\,e^{i\phi_{ij}} f^{\dagger}_{i} f^{}_{j} + \textrm{H.c.} = \sum_{s,s'}\sum_{\bs{k}}
f^{\dagger}_{s,\bs{k}} (h_{\bs{k}})_{s,s'} f^{}_{s',\bs{k}}
$, where $t=-1$ and the phase of the complex hopping {$e^{i\phi_{ij}} = e^{\pm i\,\pi/4}$ }
is depicted in Figure~\ref{fig: kagome+cartoon} (b). 
$f^{\dagger}_{s,\bs{k}}$ creates a Bloch electron with momentum $\bs{k}$ at the sublattice $s =$A, B, C. The lowest energy band of $h_{\bs{k}}$ has Chern number $1$ \cite{ohgushi2000spin, green2010isolated} and we focus on {partial} filling of this topological band, {$\nu = N^{f}/N_{u.c.} <1$}, where 
$N^{f}$ is the fermion number, and
$N_{u.c.} = N_{1} \times N_{2}$ is the number of unit cells for a torus 
extended along the primitive vectors $\bs{a}_1$ and $\bs{a}_2$.

With nearest-neighbor repulsive interactions $V_{\text{FF}} > 0$, the electronic Hamiltonian in the flat-band limit \cite{Neupert-2011,Sheng-2011,Tang-2011,Sun-2011,Regnault2011,LIU2024515}
reads
\begin{equation}
\label{eq: fermion-fermion interaction}
H_{F} = \frac{1}{2}V_{\text{FF}}\,\sum_{\langle i,j\rangle} \bar{n}^{f}_{i}\,\bar{n}^{f}_{j}    
\,,
\end{equation} 
{where $\bar{n}^{f}_{i} \equiv \bar{n}^{f}_{s,\bs{r}}$ is the electronic density projected to the lowest Chern band~\cite{parameswaran2013fractional}, with $i$ labeling sublattice $s$ and unit cell $\bs{r}$:}
\begin{equation}
\bar{n}^{f}_{s,\bs{r}} 
=\frac{1}{N_{u.c.}}
\sum_{\bs{k},\bs{k'}}e^{-i(\bs{k}-\bs{k}')\cdot\bs{r}} \langle\chi_{\bs{k}} |\boldsymbol{k},s\rangle\langle \boldsymbol{k}',s|
\chi_{\bs{k}'}\rangle
\,c^{\dagger}_{\bs{k}} c_{\bs{k}'}
\end{equation}
{with $c^{\dagger}_{\bs{k}}$ and $\ket{\chi_{\bs{k}}}$ the fermionic creation operator and Bloch eigenstate of the lowest Chern band, respectively.}

{Fig.~\ref{fig: FCI spectrum}(a) shows the spectrum of Hamiltonian \eqref{eq: fermion-fermion interaction} at filling $\nu=1/3$ for $N_f = 5$ fermions on a $3 \times 5$ lattice, supporting three quasi-degenerate ground states; the inset displays the spectral flow under twisted boundary conditions $k_1\rightarrow k_1 +\theta_x/N_1$. We define the FCI gap as the smallest energy difference between the FCI manifold and the lowest-excited states, $\Delta_\text{FCI}/V_\text{FF}=0.064$. Fig.~\ref{fig: FCI spectrum}(b) shows the quasihole spectrum on a $4\times4$ lattice, in remarkable agreement with the established FCI--FQH mapping \cite{wu2012zoology,liu2015characterization,liu2013fractional,bernevig2012emergent}: adding one flux quantum is the lattice analogue of flux insertion through a torus, generating a quasihole manifold that is nearly momentum-independent and separated from the rest of the spectrum by a clear gap. The unity-aspect-ratio lattice strongly suppresses finite-size effects from residual quasihole dispersion.} 

\begin{figure}[!t]
    \centering
    \includegraphics[width = \linewidth]{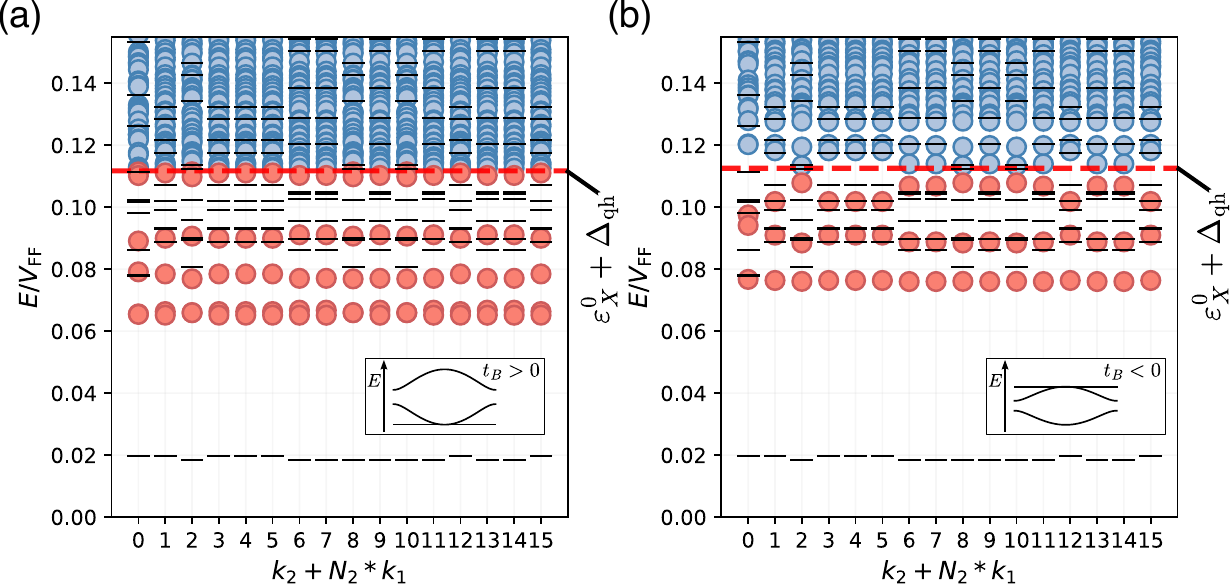}
\caption{Exciton-quasihole interacting spectra (blue and red circles) on a 4x4 lattice with $V_{\text{FB}}/V_{\text{FF}}=0.4$ and (a) $ t_\text{B}/V_{\text{FF}} = 0.12$; (b) $t_\text{B}/V_{\text{FF}} = -0.02$. Red dashed line denotes $\varepsilon_X^0+\Delta_{\text{qh}}$, below which exciton-quasihole states are characterized by negative binding energies (red circles). The quasihole spectrum is plotted in the background for reference (black dashes). Insets show the effect of opposite signs of $t_\text{B}$ on boson single-particle dispersion.}
\label{fig: quasihole spectrum}
\end{figure}

The interaction between electrons
and the exciton is described by the Hamiltonian
\begin{subequations}
\label{eq: FB total Hamiltonian}
\begin{equation}
H = H_{F} + H_{B} + H_{FB}
\,.
\end{equation}

\begin{equation}
H_{B} = 
t_\text{B}\sum_{\langle i, j \rangle}
(b^{\dagger}_{i} b_{j} + \textrm{H.c.}) + \varepsilon_0
\label{H_B}
\end{equation}
{describes nearest-neighbor boson hopping with hard-core constraint $n^{b}_{i} = b^{\dagger}_{i}\,b_{i} \in \{0,1\}$, and $\varepsilon_0$ shifts the lowest single-particle boson energy to zero ($\varepsilon_0 = 2t_\text{B}$ for $t_\text{B}>0$; $\varepsilon_0 = -4t_\text{B}$ for $t_\text{B}<0$). Unlike in the FQH-sphere setting~\cite{mostaan2025anyon,wagner2025sensing}, we treat the exciton as a mobile excitation on a toroidal lattice, preserving translation symmetry and avoiding the numerical overhead of exciton pinning. The bosonic spectrum of~\eqref{H_B} consists of a flat band and two dispersive bands~\cite{bergman2008band}: for $t_\text{B}>0$ the lowest band is flat (inset of Fig.~\ref{fig: quasihole spectrum}(a)), effectively pinning the exciton without breaking translation invariance, whereas for $t_\text{B}<0$ it is parabolic (inset of Fig.~\ref{fig: quasihole spectrum}(b)), with $t_\text{B}$ tuning the kinetic cost of binding. Tuning between these two kinematic regimes provides a key control parameter for the onset of exciton--anyon binding.} 

{
Finally, according to Fig. \ref{fig: kagome+cartoon}, the dipole interaction between the exciton and an electron separated by distance $\sqrt{r^2 + d^2}$ is
$
V_{\text{FB}}(r) = V_{\text{FB}}\left[1+(r/d)^2\right]^{-3/2}
$, where $V_{\text{FB}} = e p/(4\pi\epsilon\,d^2)$ is the onsite interaction. Due the fast power-law decay, we model the interaction as}
\begin{equation}
\label{eq: FB interaction}
{
H_{FB} = 
\sum_{i}V_{\text{FB}}\bar{n}^{f}_{i}n^{b}_{i}+
\sum_{\langle ij\rangle}
V_{\text{FB}}^{nn}\bar{n}^{f}_{i}n^{b}_{j}+
\sum_{\langle\!\langle ij\rangle\!\rangle}
V_{\text{FB}}^{nnn}\bar{n}^{f}_{i}n^{b}_{j}
\,,
}
\end{equation}
\end{subequations}
{
where
$V_{\text{FB}}, V_{\text{FB}}^{nn}, V_{\text{FB}}^{nnn} > 0$,
are, respectively, the onsite, nearest-neighbor (NN), and next-nearest-neighbor (NNN) strengths of the dipole interactions, evaluated from $V_{\textrm{FB}}(r)$ at $r=0$, $a/2$, and $\sqrt{3}a/2$, shown in Fig. \ref{fig: kagome+cartoon} (b).
}

{
For the scenario illustrated in Fig.~\ref{fig: kagome+cartoon}~(a), an FCI layer such as twisted bilayer MoTe$_2$ is placed in proximity to a TMD heterobilayer, e.g., MoSe$_2$/WSe$_2$. The relevant interaction scales can be estimated as
$V_{\text{FF}} \sim e^2/(4\pi\epsilon\,a_M)$, where $a_M$ is the moir\'e length scale of twisted bilayer MoTe$_2$, and
$V_{\text{FB}} \sim e p/(4\pi\epsilon\,d^2)$, where $p = e\,\ell$ is the dipole moment of the interlayer exciton. 
Here $\ell \approx 0.6$~nm is the charge displacement within the interlayer exciton \cite{jauregui2019electrical,shanks2021nanoscale}, in close agreement with the interlayer spacing of the heterobilayer, $\sim 0.65$~nm \cite{karni2019infrared}, and $d$ denotes the separation between the heterobilayer and the MoTe$_2$ system.
Using representative values $a_M \approx 6.8$~nm (corresponding to a $3^\circ$ twist angle) and an effective dielectric constant $\epsilon/\epsilon_0 \approx 10$ \cite{laturia2018dielectric,kutrowska2022exploring}, we obtain
$V_{\text{FF}} \sim 21.2$~meV and
$V_{\text{FB}} \sim 0.9$--$21.6$~meV for interlayer separations $d \sim 2$--$10$~nm.
Moreover, related Fermi--Bose systems have recently been realized experimentally in TMD heterobilayers such as WSe$_2$/WS$_2$, where excitons and electrons coexist within the same structure and exhibit strong electron--exciton interactions comparable to the bare electronic interaction strength \cite{xiong2023correlated,park2023dipole,gao2024excitonic,lian2024valley,wang2023intercell}. These platforms provide an alternative realization of our model and demonstrate that 
strong exciton-electron coupling
is experimentally accessible. Motivated by this, we treat $V_{\text{FF}}$ and $V_{\text{FB}}$ as phenomenological parameters and explore both weak- and strong-coupling regimes by varying the ratio $0 \leq V_{\text{FB}}/V_{\text{FF}} \leq 1$.
Henceforth, we measure all energy scales in units of the fermionic interaction $V_{\text{FF}}$.
}

\noindent
\textit{Results from Exact Diagonalization --}
{Hamiltonian \eqref{eq: FB total Hamiltonian} is translation invariant and conserves fermion and boson numbers separately. Many-body states of $N^{f}$ fermions and one boson with total momentum $\bs{Q} = \bs{p} +\sum^{N^{f}}_{i = 1}\bs{k}_i$ (mod. $\bs{G}$) take the form}
\begin{equation}
\label{eq: Q state}
\ket{\Psi^{N_f}_{\bs Q}}
=
\sum_{\{\bs k\},\bs p,s}
\Psi_{\{\bs k\},\bs p,s}\,
\Big(\prod_{i=1}^{N_f}
c^\dagger_{\bs k_i}\Big)\ket{0_f}
\otimes
b^\dagger_{\bs p, s}\ket{0_b}\, ,
\end{equation}
{where $\Psi_{\{\bs k\},\bs p,s}$ is the complex amplitude of the tensor product of a Slater determinant $\Big(\prod_{i=1}^{N_f} c^\dagger_{\bs k_i}\Big)\ket{0_f}$ and a single boson state $b^{\dagger}_{\bs{p},s}\ket{0_{b}}$ with momentum $\bs{p}$ on sublattice $s$. The Hilbert space dimension per momentum sector grows as $\dim(\mathcal{H}_{\bs{Q}})\sim \binom{N_{u.c.}}{N_f}\binom{3N_{u.c.}}{N_b}/N_{u.c.}$ (with $N_b=1$), posing substantial computational demands.} 
{For the system size $N_1\times N_2=4\times4$ considered below, this yields $\dim(\mathcal H_{\mathbf Q})\sim 10^4$. To access larger quasihole geometries while maintaining near-unit aspect ratio, we additionally project the boson onto its lowest band, which substantially reduces the Hilbert-space growth and enables calculations up to the $5\times5$ lattice, where the projected problem still reaches $\sim 10^6$ states.}

{To probe anyon--exciton binding, we track the spectral evolution as a quasihole is created both with and without an exciton, as shown in Fig.~\ref{fig: kagome+cartoon}(c). Without a boson (bottom panel), holding $N^{f}$ fixed, a single quasihole is nucleated by changing the filling from $\nu = 1/3$ to $\nu = 1/3 - \delta$; for $N^{f}=5$ electrons this is realized by adding one unit cell, from a $3 \times 5$ to a $4 \times 4$ lattice. The resulting ground-state energy difference defines the quasihole energy $\Delta_{\textrm{qh}}$. With an exciton present (top panel), we track the shift in ground-state energy from $\varepsilon_X^{0}$ -- the boson energy at $\nu=1/3$ -- to $\varepsilon_X$, the energy when the boson interacts with a state containing a quasihole, again transitioning from a $3\times5$ to a $4\times4$ lattice at fixed $N^{f}=5$.
}

The spectrum of the 
hybrid system is plotted
in Fig.~\ref{fig: quasihole spectrum}  (circles) as a function of the total momentum, and is compared with the spectrum of an isolated quasihole (dashes), identical to that shown in Fig.~\ref{fig: FCI spectrum}(b). By tuning the sign of the boson hopping, we access two distinct regimes, illustrated in Fig.~\ref{fig: quasihole spectrum}(a) for $t_\text{B}>0$ and Fig.~\ref{fig: quasihole spectrum}(b) for $t_\text{B}<0$. As highlighted in the insets, positive 
hopping yields a lowest-energy flat-band dispersion, whereas for negative hopping the low-energy bosonic states are concentrated near the minimum of the lowest parabolic band.

{A characteristic feature of Figs.~\ref{fig: quasihole spectrum}(a) and (b) is that the repulsive exciton-electron interaction raises the exciton-quasihole energies above the independent quasihole branch. In particular, we observe mid-gap states lying between the spectral minimum ($E/V_{\textrm{FF}}\approx 0.02$) and the bottom of the continuum quasihole spectrum ($E/V_{\textrm{FF}} \approx 0.08$). We highlight by a dotted red line the energy scale $\varepsilon^{0}_{X} + \Delta_{\textrm{qh}}$, the sum of the boson-electron ground state energy at $\nu=1/3$ and the quasihole energy, which defines the boundary between exciton-quasihole bound and scattering states. Remarkably, the ground state of the interacting system satisfies $\varepsilon_{X} < \varepsilon^{0}_{X} + \Delta_{\textrm{qh}}$ -- a spectral signature of anyon-exciton bound states energetically favored over the asymptotically separated quasihole and exciton. Physically, the states below the scattering continuum $\varepsilon^{0}_{X} + \Delta_{\textrm{qh}}$ -- denoted by red circles in Figs.~\ref{fig: quasihole spectrum}(a) and (b) -- arise from the net residual attraction between the exciton and the electronic charge depletion caused by the quasihole, highlighted in Fig.~\ref{fig: kagome+cartoon}(a).
} 

{
{To quantify bound-state formation and the pronounced influence of the exciton dispersion evident in Figs.~\ref{fig: quasihole spectrum}(a) and (b), we define the binding energy $\Delta\varepsilon$ as the change in ground-state energy induced by an exciton between the two configurations shown in Fig.~\ref{fig: kagome+cartoon}(c), measured relative to an isolated quasihole excitation:}
\begin{equation}
    \Delta\varepsilon = \varepsilon_X-(\varepsilon_X^0+\Delta_{\text{qh}})
\label{binding potential}
\end{equation}
{Numerically, $\Delta_{\text{qh}}$ is found to be nearly zero, reflecting both finite-size effects and the well-known tendency of the quasihole gap to be significantly smaller than the charge--neutral gap~\cite{chakraborty1986elementary,haldane1985finitesize,beran1991interaction,morf1986monte}.}
Importantly, taking $\Delta_{\text{qh}}\approx0$ yields a conservative estimate of the binding energy, 
providing
a lower bound on $|\Delta\varepsilon|$ in \eqref{binding potential}. Moreover, the characteristic binding energies we extract are sizable on the scale of 
$\Delta_{\text{FCI}}$; for example,
$|\Delta\varepsilon|/\Delta_{\text{FCI}}\approx0.73$ in Fig.\ref{fig: quasihole spectrum} (a) for moderate interaction strength $V_\text{FB}/V_\text{FF}=0.4$, {indicating that the exciton-quasihole binding 
phenomenon
is robust despite finite size effects.}

}

\begin{figure}
    \centering
    \includegraphics[width = \linewidth]{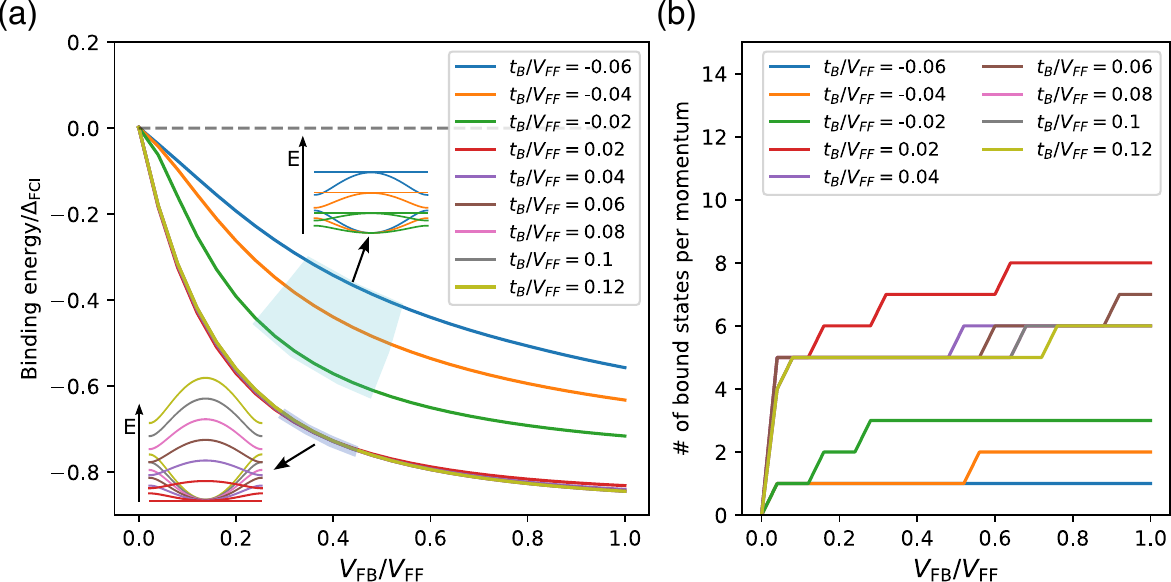}
\caption{Characterization of the exciton-quasihole binding {on a $4 \times 4$ lattice}. The binding energy $\Delta\varepsilon$ in units of FCI gap (a) and {the number of bound states per momentum sector} (b),
versus the fermion-boson interaction strength. 
Legend denotes the boson hopping strength. Inset shows the change of exciton single-particle dispersion by varying $t_\text{B}$, for positive and negative hopping.}
\label{fig: quasihole parameter}
\end{figure} 

\begin{figure}[!htb]
    \centering
    \includegraphics[width = \linewidth]{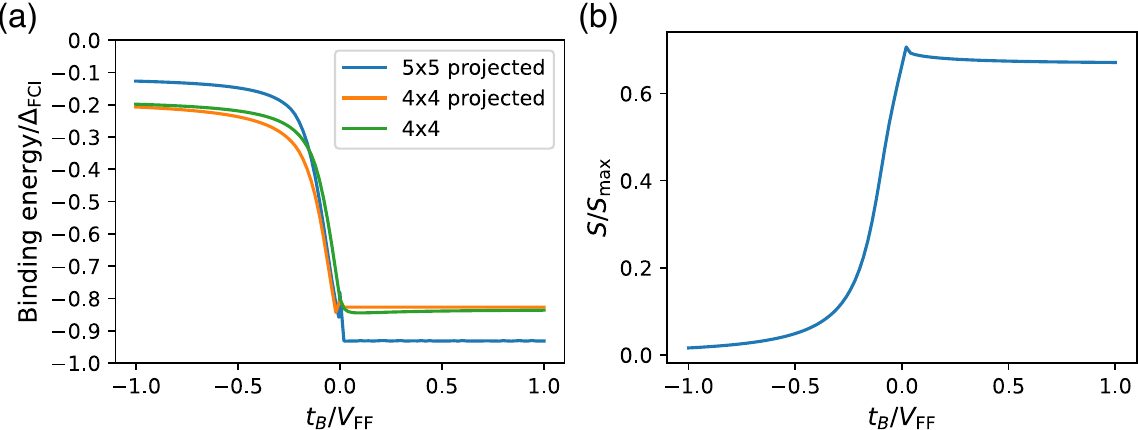}
\caption{Kinetic effects on exciton--quasihole binding at strong coupling, $V_{\text{FB}}/V_{\text{FF}}=1$. 
(a) Binding energy versus $t_\text{B}$ for the $4\times4$ lattice and for band-projected boson on the $4\times4$ and $5\times5$ lattices. 
(b) Ground-state {exciton-anyon} entanglement entropy on the $4\times4$ lattice, normalized by $S_{\max}=\log(3N_{\mathrm{u.c.}})$.
}
\label{fig: binding energy scaling and entropy}
\end{figure}

{The exciton--quasihole binding energy and {the number of bound states per center--of--mass momentum} as functions of the fermion--boson interaction strength $V_{\textrm{FB}}$ and the boson mobility $t_{\textrm{B}}$ are shown in Fig.~\ref{fig: quasihole parameter}(a) and (b), respectively. Stronger interactions and reduced boson mobility clearly enhance bound--state formation, as evidenced by the increasing binding energy and the growing number of bound states, reflecting the competition between interaction--driven binding and the boson kinetic energy. 
}

{
{
The formation of exciton--quasihole bound states exhibits a pronounced dependence on the sign of $t_{\text{B}}$. For $t_{\text{B}}>0$, the lowest boson band is exactly flat (zero group velocity, infinite effective mass), so variations in hopping strength incur a negligible kinetic penalty for binding. Consequently, the binding energy is governed primarily by $V_{\text{FB}}$ and depends weakly on $t_{\text{B}}$ (inset of Fig.~\ref{fig: quasihole parameter}(a)), while the number of bound states depends strongly on interaction strength (Fig.~\ref{fig: quasihole parameter}(b)), indicating that quasiholes bind efficiently to ``heavy'' excitons. The kagome lattice naturally realizes such heavy bosons without translation--symmetry--breaking pinning potentials, avoiding prohibitive computational costs.
}
{By contrast, for $t_{\text{B}}<0$ the lowest boson band becomes dispersive, and varying $t_{\text{B}}$ modifies the curvature near the band minimum (inset of Fig.~\ref{fig: quasihole parameter}(a)). In this regime, $t_{\text{B}}$ directly tunes the effective mass and hence the kinetic cost of binding: a lighter (i.e., more dispersive) boson is more difficult to bind to the anyon, reducing the magnitude of the binding energy, whereas a heavier (i.e., less dispersive) boson enhances binding. This provides a physical explanation for the pronounced $|t_{\text{B}}|$ dependence of the binding energy observed for $t_{\text{B}}<0$ in Fig.~\ref{fig: quasihole parameter}(a), together with the reduced number of bound states in Fig.~\ref{fig: quasihole parameter}(b), in sharp contrast to the flat--band regime ($t_{\text{B}}>0$).}

{To further assess effects of exciton kinetic energy and system size on the binding process, we analyze larger systems of sizes $4 \times 6$ and $5 \times 5$ with $N^{f}=8$ fermions. These geometries maintain near-unit aspect ratio while supporting, respectively, a three-fold FCI ground-state degeneracy and a quasi-degenerate quasihole manifold, consistent with Fig.~\ref{fig: FCI spectrum}. Projection of the boson onto its lowest band is employed to mitigate the steep increase in Hilbert space dimensionality, reaching $\sim 10^6$ states (two orders of magnitude larger than the $4 \times 4$ system).
Fig.~\ref{fig: binding energy scaling and entropy}(a) displays the binding energy versus $t_B$ at strong coupling for the $4 \times 4$ and $5 \times 5$ lattices. Importantly, boson projection does not significantly affect the binding energies: the projected $4\times4$ curve agrees closely with the full $4\times4$ result in both the $t_B>0$ and $t_B<0$ limits, corresponding to the flat-band and parabolic regimes, respectively.
For $t_B>0$, the binding energy remains stable upon increasing the system size, providing strong evidence for robust exciton--quasihole binding  when exciton kinetic energy is quenched.
}

{By contrast, kinetic effects are more pronounced for $t_B < 0$. For sufficiently negative $t_B$, the binding energy decreases systematically with system size, consistent with asymptotic decoupling of the exciton and the quasihole. This behavior is mirrored in the ground-state exciton-anyon entanglement entropy,
$S = -\mathrm{Tr}(\rho\,\log \rho)$, 
computed from the reduced density matrix $\rho = \mathrm{Tr}_{F}(\ket{\Psi_{\bs{Q}}}\bra{\Psi_{\bs{Q}}})$ after tracing out the electrons. As shown in Fig.~5(b), $S \to 0$ for large negative $t_B$, consistent with a product state $\ket{\Psi_{\bs{Q}}} \rightarrow |\mathrm{quasihole},\bs{Q}\rangle \otimes |\boldsymbol{p}=0\rangle$, whereas for $t_B > 0$ the exciton--quasihole state remains significantly entangled, saturating at  about $S_{\textrm{sat}} \approx 0.67\,S_{\max}$ 
of the maximum entanglement $S_{\max} = \log(3\,N_{\textrm{u.c.}})$.
We identify a crossover scale $t^{*} \approx -0.1\,V_{\mathrm{FF}} \approx -2.1\,\mathrm{meV}$, where the entanglement and binding energies drop by $50\%$.
These results show that binding is enhanced when the boson kinetic energy is quenched, consistent with the picture that a weakly dispersive, effectively pinned boson binds more strongly to the quasihole. Notably, this trend is qualitatively consistent with a recent experiment~\cite{li2026signatures} on anyon--trion binding in moir\'e FCI systems, where binding to fractionalized quasiparticles is favored when the trion is effectively localized.}

}

\noindent
\textit{Discussion--}
{ In this work, we investigated the formation of exciton--quasihole bound states in a kagome--lattice fermion--boson model composed of a fractional Chern insulator at $\nu=1/3$ interacting with a mobile exciton whose dispersion is tunable from parabolic to flat. Using exact diagonalization, 
we demonstrated the formation of exciton--quasihole bound states, as a function of the repulsive electron--exciton interaction $V_{\mathrm{FB}}$ and the exciton kinetic energy $t_{\mathrm{B}}$, appearing as low--lying levels clearly separated from the scattering continuum, whose binding channel emerges despite repulsive electron--exciton interactions owing to a residual attraction to the quasihole charge depletion. Reducing $t_{\mathrm{B}}$ enhances this effect by amplifying the role of interactions relative to kinetic energy.
}

{
Moir\'e transition--metal dichalcogenides provide a natural experimental platform for realizing these exciton--anyon bound states, as they host FCI states and support robust, optically active excitons. In this context, the mobile boson can be interpreted as a dipolar exciton, with $t_{\mathrm{B}}$ encoding its effective mass and $V_{\mathrm{FB}}$ setting the electron--exciton interaction scale~\cite{xiong2023correlated,park2023dipole,gao2024excitonic,lian2024valley,wang2023intercell}. The energy scales extracted from Fig.~\ref{fig: quasihole parameter}(a) indicate that weakly dispersive excitons can bind to quasiholes with energies reaching a significant fraction of the FCI gap. In particular, at strong coupling $V_{\text{FB}}/V_{\text{FF}} \sim 1$, the binding energy satisfies {$|\Delta \varepsilon|/\Delta_{\text{FCI}} \sim 0.55$--$0.85$, corresponding to $0.8$--$1.2$ meV} for twisted bilayer MoTe$_2$ parameters~($V_\text{FF}\sim 21.2$ meV). 
This suggests that exciton--anyon bound states, well separated from the scattering continuum, could persist up to experimentally relevant temperatures in heterostructures where an FCI lies within a few nanometers of an excitonic heterobilayer. Although our calculations employ periodic boundary conditions to reduce the Hilbert space, we expect that, in a picture where the exciton acts as a local pinning potential, exciton--quasihole binding will manifest as a redshift of the exciton energy, providing a direct optical signature of the quasihole. Notably, the binding scales we find are comparable to those reported in graphene FQH systems at $B=16$ T~\cite{mostaan2025anyon}, albeit here realized at zero external magnetic field.
}

{
This work opens several promising directions. First, extending the exciton--anyon framework to the fractional electronic states of twisted bilayer MoTe$_2$~\cite{lu-santos-2024fractional,lu2025electromagnetic,wang2024fractional,wang2023topology,jia2024moire,Xu2025multiple,wang2025higher,pichler2025single,liu2025characterization,yan2025anyon} will bring the analysis closer to experiment and enable a systematic assessment of how the quantum geometry of the Chern bands shapes the relevant energy scales~\cite{wang2021exact,ledwith2023vortexability,onishi2025quantum,yu2025quantum}. Second, coupling excitons to non--Abelian FCI anyons~\cite{reddy2024non,Ahn2024non-abelian,chen2025robust,liu2025non} could provide an optically addressable interface to topological qubits. Third, exploring finite exciton densities will allow investigation of the interplay between exciton--exciton interactions and the collective excitations of the FCI, opening access to highly entangled fermion--boson hybrid states.
}

\vspace{0.5cm}

\noindent
\textit{Acknowledgments --} We are grateful to 
Andrei Bernevig, 
Yves Kwan, 
Elaine Li,
Leo Li, 
Nicolas Regnault, 
and
Xiaodong Xu,  
for stimulating discussions. This research was supported by the U.S. Department of Energy, Office of Science, Basic Energy Sciences, under Award DE-SC0023327. 
This work was performed in part at Aspen Center for Physics, which is supported by National Science Foundation grant PHY-2210452.
\bibliography{bibliography}

\end{document}